\documentclass[prb, twocolumn]{revtex4-2}
\usepackage{graphicx}
\usepackage{dcolumn}
\usepackage{bm}
\usepackage [latin1]{inputenc}
\usepackage{epstopdf}
\usepackage{color}
\usepackage{ulem}
\usepackage{multirow}

\begin{document}
\title{Study of magnetoelastic interaction in MnF$_2$ by the acoustoelectric transformation method}

\author{I.\,V.\,Bilych$^1$, K.\,R.\,Zhekov$^1$, T.\,N.\,Haidamak$^{1**}$, G.\,A.\,Zvyagina$^1$, D.\,V.\,Fil$^{2,3}$, V.\,D.\,Fil$^1$}

\email{dmitriifil@gmail.com; fil@ilt.kharkov.ua}

\affiliation{$^1$B.~Verkin Institute for Low Temperature Physics and Engineering, National
Academy of Sciences of Ukraine, Nauky Avenue, 47 Kharkiv 61103, Ukraine\\
$^2$Institute for Single Crystals, National Academy of Sciences of Ukraine,
60 Nauky Avenue, Kharkiv 61072, Ukraine\\
$^3$V.N. Karazin Kharkiv National University, 4 Svobody Square, Kharkiv 61022,
Ukraine}

\begin{abstract}

The mechanisms of magnetoelastic interaction in MnF$_2$ are studied using the method of acoustoelectric transformation. The temperature dependence of the piezomagnetic coupling coefficient and its anisotropy are determined in the antiferromagnetic phase. We observe a new effect consisting in the appearance of a non-diagonal component of the magnetic susceptibility tensor, which is proportional to the square of the order parameter, under the action of the shear wave. A phenomenological interpretation of the effect, which takes into account a small-angle rotation of the crystal lattice, is presented. In the paramagnetic state, the effect of acoustic deformation is reduced to the modulation of the diagonal component of the susceptibility tensor. 
\end{abstract}

\maketitle
\section{Introduction}

In this paper, the influence of elastic strains on the state of the magnetic subsystem of MnF$_2$ single crystals is investigated by acoustic methods. The motivation for our reference to this
rather well-studied compound is due to the following circumstances. Manganese fluoride together with cobalt fluoride were the first antiferromagnets in which piezomagnetism was detected Ref. \cite{1}. The measurements  \cite{1} were performed using the standard balance method of the measurement of magnetic moments. Unfortunately, in MnF$_2$, due to a very small value of the corresponding piezomagnetic modulus, it was only possible to establish the existence of the effect and to estimate its intensity, but no measurements of the temperature dependence and anisotropy were made. Later the presence of piezomagnetism in MnF$_2$ was confirmed  \cite{2} using the SQUID magnetometer, but in that experiment the characteristics mentioned above were not studied either. The reason for a relative failure in both cases was the need to detect a weak piezomagnetic response at a significant background level unrelated to the strains.

In the present work, the piezomagnetic response in MnF$_2$ is studied using the acoustoelectric transformation (AET) method. Previously the AET method was successfully used to study piezomagnetism in CoF$_2$ \cite{3}. Significant advantages of the method are the practical absence of a background signal and a flexibility in implementing various experimental geometries. As a result, the temperature dependence of the effect and its anisotropy were reliably established, and a new phenomenon, the appearance of nondiagonal components of the magnetic susceptibility tensor in antiferromagnets in the field of a displacement gradient, was observed. 

The impact of acoustic deformations on the characteristics of the magnetic subsystem in the paramagnetic state is also investigated.

Some disadvantage of the AET method is the impossibility to provide direct quantitative measurements of the intensity of the studied effects. The measurement algorithm gives only relative quantities. But this shortage can be easily overcome by comparing the measured parameters with the response of a calibrated sample.

\section{Experimental details}

The basis of the AET method for studying the piezomagnetic effect in magnetodielectrics and the description of its experimental implementation were given in detail in Ref.  \cite{3}. Here we briefly outline the features of the measurement procedure that are essential for further discussion and relevant to the specific experiment. 

A sound wave of a given polarization, introduced into the sample through an acoustic delay line (DL), propagates along one of the main directions of the elastic modules tensor. In certain situations, due to the magnetoelastic interaction, magnetization oscillations, synchronous with acoustic ones, are excited in the sample. The magnetization oscillations are transformed at the output interface into an electromagnetic field, which is registered by a polarized antenna. To study the influence of the magnetic field $\bm{H}$ on the AET process in the described experiments the geometry  $\bm{H}\parallel \bm{q}$  ($\bm{q}$ is the sound wave vector) was used.

The waist of the sound beam is assumed to be much larger than the sound wave length. Then the problem is reduced to the one-dimensional one. Two scenarios of the AET response are possible. 

In the first scenario, the vector of magnetization $\bm{m}$ (excited by the sound) is orthogonal to the wave vector $\bm{q}$. This configuration arises due to the piezomagnetic interaction, whether an external magnetic field is present or not. The magnetization induced by the external field $\bm{H}$ can also have a component which is acoustically modulated and orthogonal to $\bm{q}$. In the discussed geometry ($\bm{H}\parallel \bm{q}$), this is possible only in the presence of non-diagonal components of the magnetic susceptibility. As we will see below, this is exactly the situation in the antiferromagnetic phase of MnF$_2$. For such a scenario (let us call it a "wave" scenario) in the  harmonic approximations ($\bm{E}, \bm{m}\propto \exp(i \omega t)$ )  in the sound beam one can write the following wave equation 
\begin{equation} \label{1}
\Delta \bm{E}=-k_c^2 \varepsilon \mu \bm{E}+4 \pi i k_c \nabla\times \bm{m}
\end{equation}
(in Gauss system of units), where $k_c=\omega/c$. In the one-dimensional case under discussion, the spatial dependence is proportional to $\exp(-i \bm{q}\cdot\bm{r})$, and  in our experiments $\bm{q}$ is orthogonal to the radiating interface.

Since $\vert \bm{q}\vert \gg k_c$, the first term in the right-hand side of Eq. (\ref{1}) can be neglected. The solution of Eq. (\ref{1}) yields the electric field $\bm{E}$ tangential to the radiating surface and orthogonal to $\bm{m}$. The electromagnetic radiation arises due to the continuity of $\bm{E}$ at the boundary. In the near zone it can be approximated by a linearly polarized plane wave. The magnetic component $\bm{h}$ of this wave is collinear to $\bm{m}$. The field $\bm{h}$ is registered by a frame antenna, which can be rotated to determine the polarization of the wave. Since $\bm{h}$ and $\bm{m}$ are collinear and proportional to each other, to simplify the following discussion we identify the electromagnetic field registered by the antenna with the acoustically induced magnetization in the sample.

At  $\bm{H}\ne 0$, another scenario (magneto-dipole) is also possible. Sound perturbation modulates the diagonal component of the magnetic susceptibility, so that a collinear to $\bm{H}$ and sign-alternating component of the magnetization appears in the sample. Averaged over the sample, it creates the magnetic field outside the sample and affects the antenna. The characteristics of this field can be calculated by methods of magnetostatics. Its structure is similar to the dipole field, i.e., it has both longitudinal and transverse components. But still the amplitude of the response registered by the antenna is related by a linear dependence to the magnetization contribution induced by the sound. However, the identification of the measured response with a particular component of $\bm{m}$ is no longer possible. 

All measurements were performed in the pulse mode. The DL allows us to separate the signals under study from the parasitic leakage of the radio pulse, which excites the piezoelectric transducer. A Ge monocrystal oriented along the second-order axis was used as the DL. The latter ensured the possibility of working with the elastic strains of a given polarization.

The apparatus described in Ref. \cite{4} was used, which allows one to register simultaneously the amplitude and the phase of the electromagnetic signals at fixed frequency ($\approx 55$ MHz).

Monocrystalline MnF$_2$ samples were in the form of plane-parallel plates with the thickness  $\approx 0.3$ mm, the transverse dimensions  $\approx 3$ mm, and the C$_4$ axis orthogonal to the plane of the plates. All experiments were performed in the $\bm{q}\parallel$ C$_4$ geometry. Two samples were investigated, which showed identical results. 
The results stated below refer to the temperature range up to 200 K, including the range of the acoustic grease melting ($\approx 120$ K). Since the acoustic path "piezoelectric transducer -- DL -- sample" includes two transition grease layers, a temperature-dependent correction was introduced into the high-temperature part of the amplitude data, correcting variations in the transition damping. To obtain it, in the same experiment we measured the temperature variations in the amplitude of the signal rereflected in the DL, which also passed twice through the transient layer of the acoustic putty. At our frequencies the temperature variations of the attenuation in the germanium DL due to its smallness can be neglected \cite{5}.

\section{Results and discussion}

\subsection{The  $u_{xz}$ deformation}

Let us first recall the information necessary for further discussion on some physical characteristics of MnF$_2$. The crystal structure of MnF$_2$ belongs to the centrosymmetric tetragonal syngony, spatial group P4$_2$/mnm. The main symmetry elements are the 4th order helical axis -- C$_4$ axis (4$_z$), the sliding symmetry plane $\bm{n}$, parallel to C$_4$, and the second order axis 2$_d$, orthogonal to C$_4$ and directed along the diagonal of the basic square. Naturally, the last two elements are duplicated by the C$_4$ axis. The generally accepted canonical choice of coordinate axes is that the $z$-axis is parallel to C$_4$, the $x$-axis is orthogonal to one of the $n_x$ sliding planes.

At $T_N \approx 68$ K there is a transition to a magnetically ordered phase -- a two-sublattice uniaxial fully compensated antiferromagnet. The order parameter is the antiferromagnetism vector $\bm{L}$ oriented along the $z$-axis.

Within the framework of crystallochemical symmetry the piezomagnetic moment $\bm{m}^0$, which appears due to the partial decompensation of the antiferromagnet under the strain, is described  by the tensor relation \cite{6}: 
\begin{equation}\label{2}
m_i^0=P_{ijkl}L_j u_{kl}.
\end{equation}
Here and hereafter, the index 0 indicates the response observed in the absence of an external magnetic field. The invariance of relations (\ref{2}) with respect to the operations of crystallochemical symmetry determines the structure of the tensor  $P_{ijkl}$. It is also necessary to take into account the parity of a particular symmetry element \cite{6}. The latter is considered odd if it connects atoms in different sublattices. An odd element changes the sign of the antiferromagnetism vector in addition to the rules for converting the components of the axial vector. In the P4$_2$/mnm group the elements $n_x$ and 4$_z$ are odd. It means, for example, that when reflected in the plane $n_x$ the  $L_z$ component does not change its sign. 

The tensor structure of $P_{ijkl}$  for different spatial groups is presented in many publications (see, for example, Ref. \cite{6}), but it is given, as a rule, only for the single choice of the canonical variant of the coordinate system. The transition to another system by the tensor algebra algorithms is rather cumbersome, but one can use more simple and physically more transparent procedure. 

In the particular situation under consideration ($\bm{q}\parallel \bm{L} \parallel \bm{z}$), the relation (\ref{2}) reduces to
\begin{equation}\label{3}
m_i^0=P_{i3k3}L_z u_{kz}.
\end{equation}

Let the canonical coordinate system with the $x$-axis orthogonal to the sliding plane be chosen.  Applying the $n_x$ operation to Eq. (\ref{3}), we can see that only the $P_{1323}$ and $P_{2313}$ components remain nonzero. The presense of 2$_d$ operation, which swaps the $x$- and $y$-axes, allows to conclude that these components of the piezomagnetism tensor are equal each other. Thus, with this choice of the coordinate axes, the piezomagnetic moment is orthogonal to the vector of displacement inducing this moment.

The sign of these components remains undefined. Let us assume that it is positive. The scheme of the mutual arrangement of the displacement vectors and  excited by them piezomagnetic moments is shown in Fig. \ref{1} by solid arrows. Note that the non-equivalence of the x- and y-axes is due to the disappearance of the C$_4$ axis in the MnF$_2$ magnetic group. 

\begin{figure}
\begin{center}
\includegraphics[width=8cm]{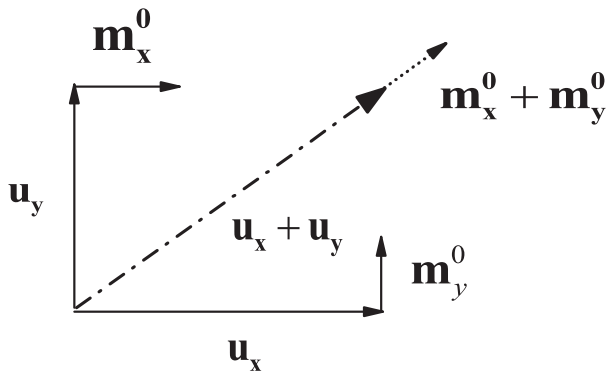}
\caption{Scheme of the piezomagnetic response  at different polarizations of the elastic field.}\label{f1}
\end{center}
\end{figure}

If the 2$_d$ axis is present in the symmetry group, this axis  can be chosen as the  $x^\prime$-axis. In such a systems of coordinates applying  C$_2$ and then  4$_z$ operation to Eq. (\ref{3}), one can easily see that only two components equal in absolute value and opposite in sign, $P^\prime_{1^\prime 3 1^\prime 3}=-P^\prime_{2^\prime 3 2^\prime 3}$, remain nonzero. Hence, if we introduce a deformation into the sample with the polarization along [110] (in the canonical system), then the excited piezomagnetic moment will be collinear to the displacement vector. A simple illustration of this statement can easily be seen from the scheme shown in Fig. \ref{f1}. The elastic displacement along [110] (the dotted line in Fig. \ref{f1}) emerges as a result of interference of two in-phase and equal in amplitude normal modes ($u_{xz}$ and $u_{yz}$). Each mode generates its own piezomagnetic moment orthogonal to it. Obviously, the resulting moment is collinear to the resulting displacement.

\begin{figure}
\begin{center}
\includegraphics[width=8cm]{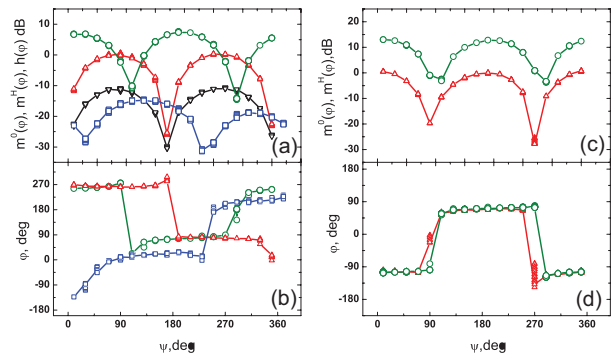}
\caption{Polarization diagrams at different polarizations of the elastic field measured in the zero magnetic field and in the presence of the magnetic field $\bm{H}$ ($\bm{H}\parallel z$): (a), (b) -- $\bm{u}\parallel [100]$, black curves corresponds to $T=53$ K, $H=0$, measurements without preliminary switching on the magnetic field; red curves corresponds to $T=53$ K, $H=0$, measurements after preliminary switching on the magnetic field $H=3$ T; green curves corresponds to $T=53$ K, $H=3$ T; blue curves corresponds to $T=72$ K, $H=3$ T;  (c), (d) -- $u\parallel [110]$,  red curves  correspond to $T=4.2$ K, $H=0$; green curves correspond to $T=4.2$ K, $H=2$ T.}\label{f2}
\end{center}
\end{figure}

Experiments have shown that in MnF$_2$ below the Neel temperature a transverse acoustic wave propagating along the C$_4$ axis always generates a piezomagnetic response. Its orientation is determined by the polarization of the elastic field. Figure \ref{f2} shows typical examples of amplitude-phase rotation diagrams observed in various experimental situations. The coordinate of the abscissa axis is the angle $\psi$ between the normal to the plane of the loop antenna and the displacement vector of the sound. Accuracy of its tuning to an initial reference level is $\pm 5^\circ$. The ordinates show the amplitude (Fig. \ref{2} (a), (c)) and the phase (Fig. 2 (b), (d)) of the recorded piezomagnetic response.

We emphasize that in Fig. \ref{f2} the coordinate corresponding to the maximum of the amplitude response determines the angle between the displacement $\bm{u}$ in the elastic deformation and the magnetic moment $\bm{m}$ excited by it. The amplitude dependences are given in semi-logarithmic scale for the sake of better visibility. The maximum signal level in the absence of a magnetic field is chosen as the reference level.   The shape of phase characteristics makes it possible to evaluate the degree of ellipticity of the registered radiation. In particular, in Fig. \ref{f2} (b), (d) the step-like phase change at the angle $\varphi \approx 180^\circ$  indicates that in these cases the polarization is practically linear.

In Fig. \ref{f2}, in addition to the dependences taken in the zero fields, the rotation diagrams measured in the presence of the magnetic field $\bm{H}$ ($\bm{H}\parallel \bm{z}$) are shown. We will turn to their interpretation later.

It follows from Fig. \ref{f2} that at $\bm{u}\parallel [100]$ (Fig. \ref{f2}(a)) $\bm{m}^0$ is orthogonal to $\bm{u}$, while at the displacement along [110] (Fig. \ref{f2}(c)) the corresponding vectors are collinear. These results are in complete agreement with the above analysis. 

It follows from Eqs. (\ref{2}) and (\ref{3}) that in antiphase domains the moments generated by the deformation are reversed, so that the resulting response can be greatly underestimated. In this connection Fig. \ref{f2} (a) illustrates a "successful" experiment of monodomainization of the sample by the short-term switching on the external field $H_z =3$ T before measurements. The result is not obvious at all, since it is necessary to specify the reason why the order parameter entered to the expansion of thermodynamic potentials in an odd degree, providing the realization of energy minimization at a given direction of $\bm{L}$. In our opinion, the piezomagnetic moment allowed by symmetry, $m_z^0\sim L_z \sigma_{xy} $, emerges in the sample rigidly glued to the DL due to the appearance of thermoelastic $\sigma_{xy}$ static stresses. Its Zeeman interaction with the external magnetic field $H_z$ leads to the monodomainization of the antiferromagnetic structure.

\begin{figure}
\begin{center}
\includegraphics[width=8cm]{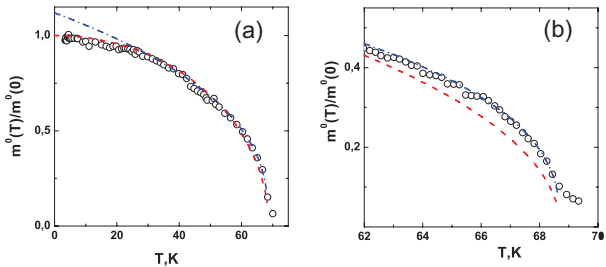}
\caption{Temperature dependence of $m^0$. Symbols are the normalized experimental values of $m^0$,  red (dash) curve is the function  $\sqrt{1-(T/68.7)^2}$,  blue (dash dot) curve is the function $1.12 (1-T/68.7)^{0.382}$. (a), the dependencies in whole temperature range; (b), the dependences in the vicinity of $T_N$.}\label{f3}
\end{center}
\end{figure}

The temperature dependence of the piezomagnetic response in the absence of an external magnetic field, measured when the antenna is tuned to the maximum signal, is shown in Fig. \ref{f3}. 

On average, it is satisfactorily described by the dependence $m^0(T)/m^0(0)\approx
\sqrt{1-(T/T_N)^2}$ (typical to the type II phase transitions), although it is not good enough near $T_N$. We will return to the discussion of the behavior in the region of critical fluctuations later. The absolute value of the piezomagnetic modulus $\lambda_{123}=P_{1z23}L_{z0}=4\pm 1$ G ($L_{z0}$
is the upper limit value of the order parameter) is determined from a comparison of the response from MnF$_2$ with the previously measured response from CoF$_2$ ($1000$  G \cite{3}). It is close to one found from the estimate presented in \cite{1} ($ 6.4$ G).

\begin{figure}
\begin{center}
\includegraphics[width=8cm]{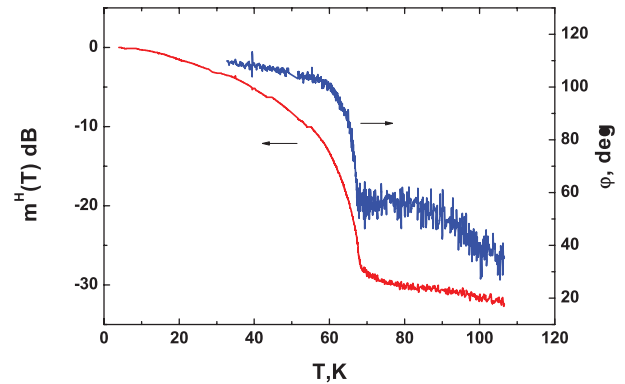}
\caption{Temperature dependence of the amplitude (red line) and the phase (blue line) of $m^H$  ($H=3$ T).}\label{f4}
\end{center}
\end{figure}

Let us now discuss the nature of the AET response $\bm{m}^H$ observed in the magnetic field, which in the case of $\bm{u}\parallel [100]$ has a polarization diagram different from that of $\bm{m}^0$. It follows from the rotation diagrams presented in Fig. \ref{f2} that its orientation is always almost collinear to the elastic displacement. In particular, at $\bm{u}\parallel [100]$, when $\bm{m}^0$ is orthogonal to the displacement, one can fix the angular coordinates at which only the $\bm{m}^H$ signal is mostly registered. The temperature behavior of the amplitude and phase, measured in such geometry, is shown in Fig. \ref{f4}. The amplitude is given in the semi-logarithmic scale, which allows to demonstrate more clearly the existence of the response even at $T > T_N$. The phase at $T_N$ changes almost in a step-like manner. 

Our interpretation of the dependences presented in Fig. \ref{f4} is as follows. The response $\bm{m}^H$ is the result of the interference of two contributions whose phases are different and are almost independent of the temperature. The first one, rather weak, apparently exists at temperatures both above and below $T_N$, while the second one, with rapidly increasing intensity, appears only in the antiferromagnetic phase. 

\begin{figure}
\begin{center}
\includegraphics[width=8cm]{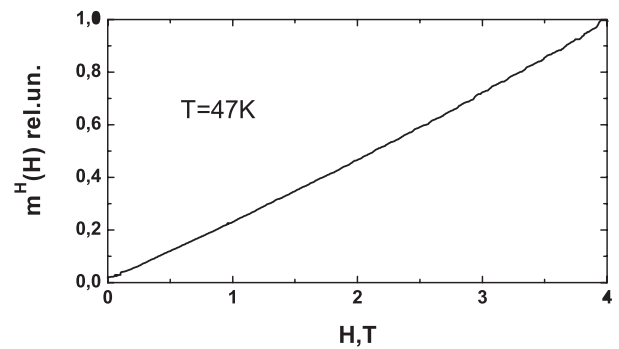}
\caption{Magnetic field dependence of $m^H$ ($T=47$ K).}\label{f5}
\end{center}
\end{figure}

Figure \ref{f5} shows the dependence of $m_x^H$ on the magnetic field, measured at the orientation of the receiving antenna, which provides the minimum piezomagnetic response $\bf{m}^0$. It is practically linear. 

Within the scenarios described in the previous subsection, the following variants are possible to explain the origin of $\bm{m}^H$. \textit{The "wave" scenario.} The deformation initiates the emergence of a nondiagonal component in the magnetic susceptibility tensor, i.e., a magnetization component orthogonal to $\bm{H}$ appears. From the formal point of view, this effect is quite possible because the field of shear displacement gradients changes the symmetry of the lattice, making it more anisotropic. \textit{The "magneto-dipole" scenario.} The deformation modulates the longitudinal relative to $\bm{H}$ magnetization, and the antenna only registers the transverse component of the magnetic field induced by magnetization. In our opinion, this scenario is unlikely. It should be expected that in this case the recorded response should be correlated with the behavior of the longitudinal magnetic susceptibility. In antiferromagnets MnF$_2$, the longitudinal magnetic susceptibility decreases with decreasing temperature, while $m_x^H$  only increases below $T_N$ (Fig. \ref{f4}).

According to the general theory of type II phase transitions \cite{7}, the temperature behavior of thermodynamic characteristics in the region of critical fluctuations is described by power law dependences on the quantity $(1-T/T_c)$, and the exponents are called critical indices. Their numerical values are universal: they depend mostly on the system dimensionality. In particular, a critical index $\beta\approx 0.33$ is expected for the order parameter in the three-dimensional case of the antiferromagnet. Experimentally, such behavior is carried out   only approximately, and the deviation of values of critical indices from theoretical ones is about $10-20$\%.

\begin{figure}
\begin{center}
\includegraphics[width=8cm]{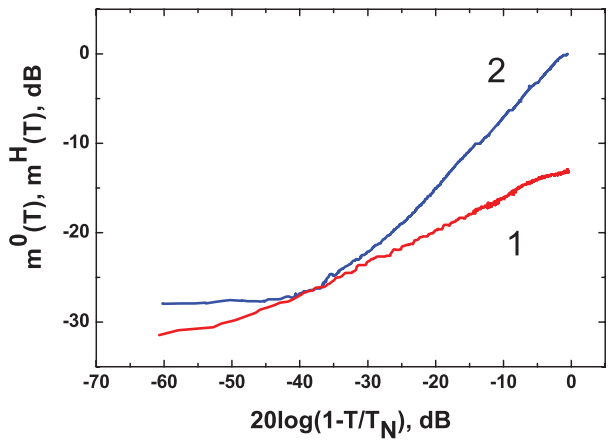}
\caption{Approximations of the temperature dependences by critical exponents. Curve 1  is $m^0$, curve 2 is $m^H$ ($H=3$ T).}\label{f6}
\end{center}
\end{figure}

\begin{figure}
\begin{center}
\includegraphics[width=8cm]{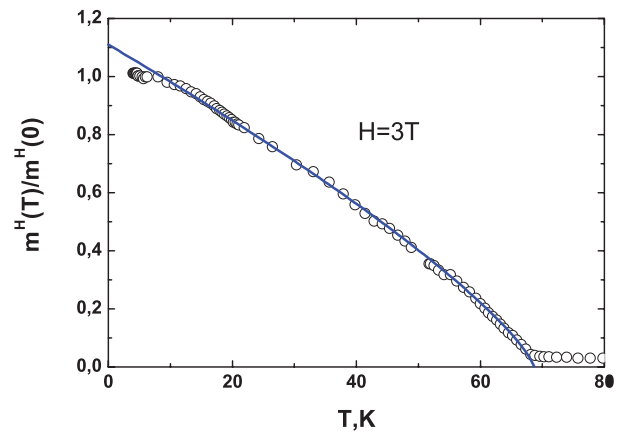}
\caption{Approximation of the normalized experimental temperature dependence of $m_x^H$ (symbols) at $H=3$ T by the function $1.11 (1-T/68.7)^{0.78}$ (blue line).}\label{f7}
\end{center}
\end{figure}

Figure \ref{f6} shows the temperature dependence of the measured signals in double logarithmic coordinates. In these coordinates the piezomagnetic response $m^0(T)$ is almost a straight line (curve 1) in the whole temperature range. A small deviation from the straight line in the left-hand part of the figure may be caused by some smearing of the transition due to the heterogeneity of the sample. The slope of the linear dependence determines the parameter $\beta=0.38\pm 0.01$. It practically coincides with the critical index value found in Ref. \cite{8} ($0.36 \pm 0.02$), characterizing the temperature behavior of the order parameter. This conclusion is in the complete agreement with the relation (\ref{3}). The accuracy of the description of the behavior $m^0(T)$ by this critical exponent is illustrated in Fig. \ref{f3}. Concidering the maximum length linear segment of the curve 1 in Fig. \ref{f6}, we refine the value of $T_N=68.7$ K. This value is used in all evaluations in this work. 
Curve 2 in Fig. \ref{f6} illustrates the temperature dependence of the response $m_x^H$. The left-hand part of this curve deviates significantly from the straight line already at $T>66$  K due to the existence of the previously mentioned small response of a different nature at all temperatures near $T_N$. Nevertheless, the slope of the straight line section ($0.78\pm 0.01$) is twice as large as the slope of line 1. It is natural to assume that the temperature dependence of the ampliture of the response is determined by the square of the order parameter  in a certain temperature range. The accuracy of the approximation by the assumed dependence is illustrated in Fig. \ref{f7}. Surprisingly, the critical exponent describes the behavior over the amlost entire temperature range below $T_N$. Thus, the response $\bm{m}^H$ can be presented as
\begin{equation}\label{4}
m_i^H=\left( B_{i3k3}L_z^2\frac{\partial u_k}{\partial z}\right) H_z.
\end{equation}

From the symmetry considerations similar to ones used above, it is easy to see that only two  components of the tensor in Eq. (\ref{4}) can be nonzero: $B_{1313}=B_{2323}$. This conclusion is valid at any choice of the coordinate system, i.e., the response is always collinear to the displacement that initiated it, which fully corresponds to the results presented in Fig. \ref{f2}. It is natural to interpret the combination in parentheses in Eq. (\ref{4}) as a non-diagonal component of the susceptibility tensor. The presence of $L_z^2$ in Eq. (\ref{4}) means that the discussed response appears only in the magnetically ordered phase.

Let us give  a simple phenomenological interpretation of the experimentally derived relation (\ref{4}). One considers that the shear wave is accompanied by a rotation of the crystal lattice by a small angle  $\delta \varphi =0.5 \nabla \bm{u}$ \cite{9}. In the framework of the Landau's approach to the expansion of the thermodynamic potential of antiferromagnets by magnetic variables, in the main approximation the magnetization is given by  equation \cite{9}
\begin{equation}\label{5}
\bm{M}=\chi_p \bm{H}- 2 D (\bm{L}\cdot \bm{H})\bm{L},
\end{equation}
where $\chi_p\approx 10^{-3}$ is the susceptibility in the paramagnetic phase at $T=T_N$, and $D$ is a dimensional (1/G$^2$) numerical coefficient.

Along with the rotation of the anisotropy axis, the antiferromagnetism vector also rotates. As a result, a component  $L_x=L_z \delta \varphi=0.5 L_z (\partial u_x/\partial z)$ appears, and in the main approximation we have the same as Eq. (\ref{4}) equation for the magnetization induced by joint action of the magnetic field and sound, 
\begin{equation}\label{6}
m_x^H=\left( D L_z^2\frac{\partial u_x}{\partial z}\right) H_z.
\end{equation}
Let us also estimate the value of the effect predicted by Eq. (\ref{6}). Since in MnF$_2$ at $T\to 0$ the susceptibility $\chi$ goes to zero, from Eq. (\ref{5}) we obtain $DL_{z0}^2\sim 0.5 \chi_p\sim 0.5\cdot 10^{-3}$. From Fig. \ref{f6}, in which both dependences are measured at the same value of $\partial u_x/\partial z$, it can be found that at low temperature  the $m^H$ signal  (at $H=3$ T)  is in 4 times larger than the $m^0$ sygnal. Taking into account the linear dependence of $m^H$ on magnetic field (Fig. \ref{f5}) we estimate that 
$m^H$ compares with $m^0$ at $H\approx 0.7$ T. Consequently, the estimate of the piezomagnetic modulus from Eq. (\ref{6}) gives  $\lambda_{xxz}\sim 3.5$ G that agrees  with the value found above ($4$ G).
Note that the realistic value of $\delta \varphi$ is  $10^{-5}$, and apparently for this reason, as far as the authors know, no such effect has been observed in other experiments. We believe that this result nicely illustrates the capabilities of the AET technique.

Let us say a few words about  possible nature of the response registered at $T>T_N$. The multiplier $L_z^2$ does not introduce any symmetry restrictions (or simplifications), and in the paramagnetic phase one would expect a realization of the relation similar to (4), but without $L_z^2$. However, in this case, the amplitude-phase diagrams at $T>T_N$ would repeat those for the antiferromagnetic phase in the position of characteristic points and shape, which in fact does not occur (Fig. 2 (a), (b)). We believe that at $T>T_N$ the second "magneto-dipole" scenario is the case, namely, the sound deformation modulates the longitudinal susceptibility, and the antenna registers the transverse component of the dipole magnetic field. The polarization diagram in this case simply reflects the asymmetry of the measuring cell together with the sample relative to the antenna rotation axis.

\subsection{The $u_{zz}$ deformation.}

In this subsection we consider the effect of $u_{zz}$ deformation on the magnetic subsystem. Such a deformation emerges under propagating of the longitudinal wave  along the C$_4$ axis. This deformation does not violate the original symmetry, so the relations of the type (\ref{4}) with $j=k=z$ are only allowed. In this case the coefficient at $H_z$  is simply the first term of the expansion of the diagonal component of the magnetic susceptibility tensor with respect to  $u_{zz}$, and hence, the response registered by the antenna is completely within the framework of the "magneto-dipole" scenario. 

\begin{figure}
\begin{center}
\includegraphics[width=8cm]{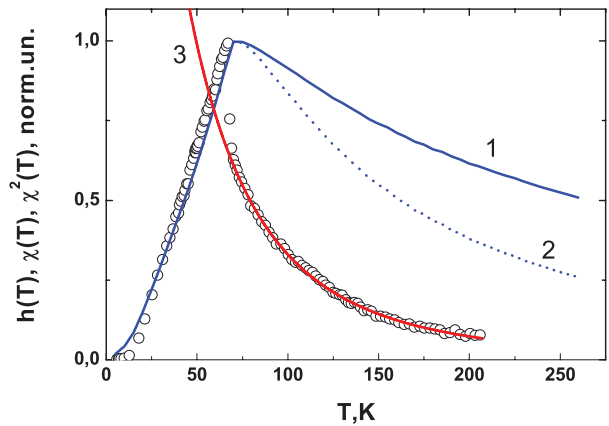}
\caption{ Temperature dependence of the AET response to $u_{zz}$ deformation and of the susceptibility. Symbols are the AET experiment, curve 1 is $\chi(T)$, curve 2 is $\chi^2(T)$ , curve 3 is the function $4.91/(1+T/86)^{3.5}$}\label{f8}
\end{center}
\end{figure}

Figure \ref{f8} shows the temperature dependence of the AET response to the  deformation in MnF$_2$  and of the magnetic susceptibility measured for the same sample. Both dependences were taken in the $\bm{H}\parallel z$ geometry and are normalized for convenience of comparison. In the antiferromagnetic state, both dependences demonstrate almost the same change, close to the linear one. The susceptibility is obviously quite well described by the Landau theory, which predicts a linear decrease in  $\chi$ with decreasing temperature \cite{9}. The AET response, which is proportional to the derivative of $\chi$ with respect to strain, should naturally exhibit a similar behavior.

Now let's turn to the paramagnetic phase. MnF$_2$ belongs to the family of classical antiferromagnets, and in the molecular field approximation its susceptibility at $T> 150$ K  is well described by the Curie-Weiss law $\chi(T)=C/(T+\Theta)$. The value of the paramagnetic Curie temperature $\Theta=86$ K was evaluated from processing of the curve 1 in Fig. \ref{f8}. This temperature is determined by the exchange interaction of given magnetic atom with the average field created by the entire ensemble of magnetic atoms. Small deviations from this law at  $T_N<T<150$ K are associated with the development of pair correlations between the nearest neighbors in the magnetic subsystem, which are not taken into account accurately by the mean-field model. 

Acoustic deformation, modulating the mutual arrangement of magnetic atoms, thereby modulates the exchange field, changing the value of $\Theta$, and with it, changing $\chi(T)$. As a result, the sample placed in a constant magnetic field acquires an additional magnetization oscillating in time and space, producing also the corresponding dipole magnetic field $h(T)\sim d \chi/d u_{zz}\sim (T+\Theta)^2 (d \Theta/d u_{zz})$. If we assume that the derivative $d \Theta/d u_{zz}$ does not depend on temperature, then the field  $h(T)$ should change as $\chi^2(T)$. In Fig. \ref{f8} for $T>T_N$ this dependence is also shown for comparison. As one can see, the behavior of $h(T)$ is noticeably different from the expected one.

\begin{figure}
\begin{center}
\includegraphics[width=8cm]{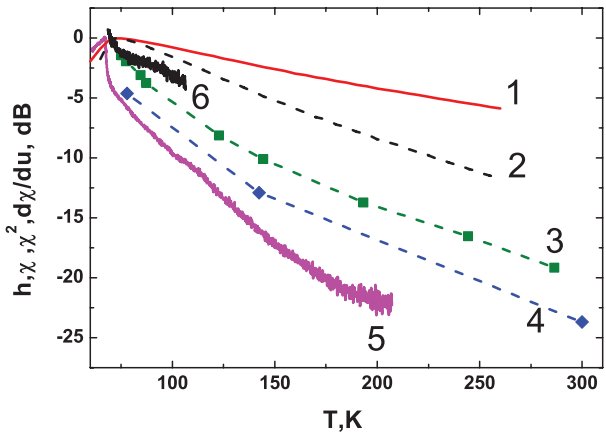}
\caption{Comparative analysis of temperature dependences of various thermodynamic functions. Curve 1 is $\chi(T)$ , curve 2 is $\chi^2(T)$ , curve 3 is $d \chi(T)/d u$ \cite{11}, curve 4 is  $d \chi(T)/d u$ \cite{10}, curve 5 is the $h(T)$ response to $u_{zz}$ (Fig.\ref{f8}), curve 6 is $m_H$ response to $u_{xz}$ (Fig. \ref{f4}). }\label{f9}
\end{center}
\end{figure}

The behavior of the MnF$_2$ susceptibility under the elastic deformation was studied by a direct method in experiments on hydrostatic compression \cite{10}. It was also estimated from the results of the study of the magnetostriction \cite{11}. Comparison of the results of this work with the literature data is shown in Fig. \ref{f9}. In this figure, in semi-logarithmic coordinates, the behavior at $T > T_N$ of the dependences $\chi(T)$, $\chi^2(T)$, $d \chi/d u$ \cite{10},  $d \chi/ d u$ \cite{11}, and our  data for $h(T)$ (from Fig.\ref{f8}) are shown. 

We emphasize that the results presented in Fig. \ref{f9} display relative temperature changes in the corresponding dependences. Their mutual relations are chosen to provide convenience of comparison. It can be seen that the results of \cite{10, 11} reach the expected proportionality $\chi^2(T)$  only at $T\geq 150$ K. It seems that the dependence $h(T)$ also tends to reach the expected behavior, although for larger certainty the temperature interval should be extended. Below 150 K, deviations from the expected dependence correspond to a more pronounced effect of deformation on pair correlations. It is especially significant in our measurements. Perhaps this is due to the frequency dependence of the correlation corrections.

It is interesting to note that the temperature change  $h(T)$ in the case under discussion is almost perfectly approximated by the power law dependence $h(T)\propto (T+\Theta)^{-3.5}$ (see Fig. 8). The deviation of the approximation from the experiment in the interval 70 K $\leq T\leq 200$ K does not exceed 5\%. It is not clear whether there is any hidden physical meaning of this behavior, or that coincidence is accidental.

In Fig. \ref{f9}, the dependence of  $h(T)$ measured in the paramagnetic phase under a shear deformation is also displayed (see the data of Fig. \ref{f4}). Its behavior seems to be proportional to $\chi^2(T)$ once $T$ exceeds $T_N$. One can think that the shear deformation, which does not change, in the first approximation, interatomic distances, has a little effect on pair correlations.

\section{Conclusion}

In summary,  the main result is that the acoustic-electrical transformation can be considered as highly informative method of the study of various mechanisms of magneto-elastic interaction in magnetoactive media. In particular, the manifestations of piezomagnetism in MnF$_2$ were studied in detail, the anisotropy of the effect and its temperature behavior were determined, and a phenomenological interpretation of the observed dependences was given. 

The novel for an antiferromagnet phenomenon, namely, the occurrence of new (off-diagonal) components of the magnetic susceptibility tensor in the field of displacement gradients, has been discovered. The form of the term of the thermodynamic potential responsible for the appearance of such an effect, which is proportional to the square of the antiferromagnetic order parameter, is experimentally established. It is shown that the effect is fully (qualitatively and quantitatively) explainable by taking into account the low-angle rotation of the lattice in the transverse acoustic wave.

 In the paramagnetic phase, AET measurements made it possible to determine the deformation contribution to the temperature behavior of the diagonal component of the magnetic susceptibility tensor, which turned out to be significant in the region of pair correlations.

\section*{ACKNOWLEDGMENT}

G.A.Z.  acknowledges the support from Pauli Ukraine Project, funded in the Wolfgang Pauli Institut Thematic Program "Mathematics-Magnetism-Materials (2021/2022)".

\vspace{1 cm}

$^{**}$\small{Present affiliation is Charles University, Prague, Czech Republic	}

\end{document}